\documentclass[12pt]{article}
\pdfoutput=1 

\usepackage{amssymb,amsmath,mathrsfs,epstopdf,slashed,color}
\usepackage{tikz}
\usetikzlibrary{matrix}
\usetikzlibrary{positioning}
\usepackage{fancyhdr} 
\usepackage{lastpage} 
\usepackage{extramarks} 
\usepackage{graphicx} 
\usepackage{lipsum} 
\usepackage{amsmath}
\usepackage{amsfonts}
\usepackage{amssymb}
\usepackage{latexsym}
\usepackage{color}
\usepackage{xypic}
\usepackage{cancel,slashed}
\usepackage[linktocpage]{hyperref}
\usepackage{booktabs}
\usepackage{caption}
\usepackage{comment}

\usepackage{graphicx}
\usepackage{placeins}
\usepackage{hyperref}
\usepackage{cite}

\newcommand{\de}{\partial}

\def\Re           {{\rm Re\hskip0.1em}}

\def\calo         {{\cal O}}

\def\calw         {{\cal W}}

\def\reals        {{\mathbb R}}

\allowdisplaybreaks[1]

\makeatletter

\@addtoreset{equation}{section}
\makeatother

\newcommand{\vol}{\mathrm{vol}}

\title{\bf Relating the Electroweak Scale to the Cosmological Constant}
\author{Stefano Andriolo, Shing Yan Li and S.-H. Henry Tye}

\begin{document}
\begin{titlepage}

\setcounter{page}{0}
  
\begin{flushright}
 \small
 \normalsize
\end{flushright}

\vskip 3cm
\begin{center}

{\Large \bf The Cosmological Constant and the Electroweak Scale}  

\vskip 2cm
  
{\large Stefano Andriolo${}^{1,2}$, Shing Yan Li${}^2$  and S.-H. Henry Tye${}^{1,2,3}$}
 
 \vskip 0.6cm

 ${}^1$ Jockey Club Institute for Advanced Study, Hong Kong University of Science and Technology, Hong Kong\\
 ${}^2$ Department of Physics, Hong Kong University of Science and Technology, Hong Kong\\
 ${}^3$ Laboratory for Elementary-Particle Physics, Cornell University, Ithaca, NY 14853, USA

 \vskip 0.4cm

Email: \href{mailto: sandriolo@connect.ust.hk, syliah@connect.ust.hk, iastye@ust.hk}{sandriolo at connect.ust.hk, syliah at connect.ust.hk, iastye at ust.hk}

\vskip 1.0cm
  
\abstract{\normalsize
String theory has no parameter except the string scale, so a dynamically compactified solution to 4 dimensional spacetime should determine both the Planck scale and the cosmological constant $\Lambda$. In the racetrack K\"ahler uplift flux compactification model in Type IIB theory, where the string theory landscape is generated by scanning over discrete values of all the flux parameters, a statistical preference for an exponentially small $\Lambda$ is found to be natural \cite{Sumitomo:2013vla}. Within this framework and matching the median $\Lambda$ value to the observed $\Lambda$, a mass scale ${\bf m}\simeq 100$ GeV naturally appears. We explain how the electroweak scale can be identified with this mass scale.
}
  
\vspace{1cm}
\begin{flushleft}
 \today
\end{flushleft}
 
\end{center}
\end{titlepage}

\setcounter{page}{1}
\setcounter{footnote}{0}

\tableofcontents

\parskip=5pt

\section{Introduction}

Cosmological data strongly indicates that our universe has a vanishingly small positive 
cosmological constant $\Lambda$  (vacuum energy density) as the dark energy, 
$\Lambda \sim 10^{-122}\ M_P^4$ \cite{Hinshaw:2012aka},
where the Planck mass  $M_P = G_N^{-1/2}\simeq 10^{19}$ GeV.  The smallness of $\Lambda$
is a major puzzle in physics.
In general relativity, $\Lambda$ is a free arbitrary parameter one can introduce, so its smallness can be accommodated but not explained within the field theory framework.\footnote{If the dark energy comes from another source, e.g., quintessence, the $\Lambda$ parameter in Einstein theory has to be even smaller.}
On the other hand, string theory has only a single parameter, namely the string scale $M_S=1/\sqrt{2 \pi \alpha' }$, so everything else should be calculable for each string theory solution. 
Since both $M_P$ and $\Lambda$ are calculable, $\Lambda$ can be determined in terms of $M_P$ dynamically in each classically stable 4-d vacuum solution. 
So we may find an explanation for a very small positive $\Lambda$. 
This happens if a good fraction of the meta-stable deSitter (dS) vacua in the relevant regions of the string landscape tend to have a very small $\Lambda$, as is the case in the racetrack K\"ahler uplift (RKU) scenario in flux compactification in string theory \cite{Sumitomo:2013vla}. Note that both the K\"ahler uplift model \cite{Balasubramanian:2004uy,Westphal:2006tn,Rummel:2011cd,deAlwis:2011dp,Sumitomo:2012vx} and the racetrack model \cite{Krasnikov:1987jj,Taylor:1990wr,Denef:2004dm} are scenarios well explored in string phenomenology.

To simplify the discussion, let us focus on flux compactification of Type IIB theory to 4 dimensional spacetime. 
Start with the four-dimensional low energy supergravity effective potential $V (F^i, \phi_j)$, where  $F^i$ are the field strengths and $\phi_j$ are the moduli (and dilaton) describing the size and shape of the compactified manifold as well as the coupling.
It is known that the field strengths $F^i$ in flux compactification in string theory take only quantized values \cite{Bousso:2000xa}, 
and all parameters like masses and couplings are now functions of discrete flux parameters $F^i$.  The string landscape is generated as we scan over all discrete values of $F^i$. That is, $V(F^i, \phi_j)$ has no free parameter, though it does contain (in principle) calculable quantities like $\alpha'$ corrections and geometric quantities like Euler index $\chi$ etc.. With no parameters to adjust, the radiative instability problem is absent \cite{Tye:2016jzi}, as ranges of flux parameters scanned over have already included the values to be fixed both before and after radiative corrections. 

For a given set of flux parameters ${F^i}$, we can solve $V(F^i, \phi_j)$ for its meta-stable vacuum solutions via finding the values $\phi_{j, {\rm min}} (F^i)$ at each solution and determine its vacuum energy density $\Lambda=\Lambda(F^i, \phi_{j, {\rm min}} (F^i))=\Lambda (F^i)$. Collecting all such solutions and feeding in a properly normalized probability distribution $P_i(F^i)$ for each $F^i$, we can determine the properly normalized probability distribution $P(\Lambda)$ of $\Lambda$ of these meta-stable solutions as we sweep through all discrete values of the flux parameters $F^i$.  
Assuming a ``dense discretuum" for each $F^i$, we may treat each $P_i(F^i)$ as a continuous function.
For smooth $P_i(F^i)$, simple probability properties show that $P(\Lambda)$ easily diverges at $\Lambda=0$ \cite{Sumitomo:2012wa}, implying that a small $\Lambda$ is statistically preferred. Ref.\cite{Sumitomo:2013vla} finds that the resulting $P(\Lambda)$ in the racetrack K\"ahler uplift scenario diverges (i.e., peaks) sharply at $\Lambda =0$. That is, an overwhelmingly large number of meta-stable vacua have an exponentially small positive $\Lambda$, so statistically, we should end up in one of them. In short, a dS vacuum with a very small $\Lambda$ is statistically natural. 

Remarkably, taking the median value $\Lambda_{50}$ in this racetrack K\"ahler uplift model to match the observed value, a natural scale emerges,
   \begin{align} \Lambda_{50}  \simeq 10^{-122}\,M_P^4 \quad  \Rightarrow \quad  {\bf m} \sim 10^2 \, {\rm GeV} \label{L2} \end{align}
Can this scale ${\bf m}$ correspond to the electroweak scale (the Higgs mass $m_H$)? In this paper, we give an explicit string theory scenario to realize this property in a concrete statistical way. Namely, we explicitly show how to introduce a Higgs-like field in the racetrack K\"aher uplift model such that $m_H\sim {\bf m}$. We argue that, in the absence of fine-tuning, statistically, the electroweak scale $m_H \sim {\bf m}$. 
This suggests that string phenomenology should focus in regions of the landscape where $\Lambda$ is naturally very small.

Notice that, following the standard supergravity formalism, the F-term effective potential $V$ and its minimum are quadratic in the superpotential $W$ and/or its derivative. So it should not be a surprise that dimensional arguments alone suggest ${\bf m} \sim  |W|^{1/3} \sim |{\Lambda}M_P^2|^{1/6}$. Also note that the emergence of the scale $\bf m$ does not entail any knowledge of the actual standard electroweak model.

\section{A Racetrack K\"ahler Uplift Model of Flux Compactification}
\label{sec:racetrack}

To be specific, let us review the racetrack K\"ahler uplift model studied in Ref\cite{Sumitomo:2013vla}, with the addition of a Higgs-like field. We consider a 6-dimensional Calabi-Yau (CY) manifold $M$ with a single ($h^{1,1}=1$) K\"ahler modulus $T$ and two (or three, i.e., $h^{2,1}=2$ or $3$) complex structure moduli $U_i$, so the manifold $M$ has Euler number $\chi(M)=2(h^{1,1}-h^{2,1}) <0$.
This simplified model of interest is motivated by orientifolded orbifolds \cite{Lust:2005dy,Lust:2006zg}, and it is given by (setting $M_P=1$),
   \begin{align}
   \label{LVS}
  &V = e^{K} \left(K^{I \bar{J}} D_I W D_{\bar{J}} {\overline W} - 3\left|W \right|^2\right),\\ \nonumber
     &K = K_{\rm K} + K_{\rm d} + K_{\rm cs} +K_H 
     = -2 \ln({\cal V} + \xi/2) -\ln(S+\bar{S}) -\sum_{i=1}^{h^{2,1}} \ln(U_i + \bar{U}_i) +K_H ,\\ \nonumber
     &W =  W_0(U_i,S,\phi) +  W_{NP}, \qquad
     W_0(U_i,S,\phi) = W_0(U_i,S) + W_0(\phi), \\ \nonumber
     &W_0(U_i,S) =  c_1 - Sc_2 +\sum_{i=1}^{h^{2,1}} (b_i - S d_i) U_i 
     , \qquad
     W_{NP} = A e^{-a T} + B e^{-bT} \,.
\end{align}
Here, ${\cal V}\equiv\vol/\alpha'^3=(T + \bar{T})^{3/2}$, $\xi\propto -\chi(M)(S + \bar{S})^{3/2}>0$, and $M_P^2 \simeq {\cal V}/\alpha'$. The two terms in the non-perturbative $W_{NP}$ form the racetrack (and stabilise $T$ \cite{Kachru:2003aw}). They are given by gaugino condensates with coefficients $a=2\pi/N_1,b=2\pi/N_2$ for $SU(N_1),SU(N_2)$ gauge symmetry respectively. The flux parameters $c_i, b_i$, $d_i$, $A$, $B$ 
are to be treated as independent (real) 
variables with smooth probability distributions that allow the zero values, while the dilation $S$ and the complex structure moduli $U_i$ are to be determined dynamically.  $W_0(\phi)$ and $K_H$ for the Higgs-like $\phi$ will be discussed later. The model also includes the first $\alpha'$-correction (the ${\xi}$ term) to the K\"ahler potential to lift the supersymmetric solution to de Sitter space \cite{Becker:2002nn,Bonetti:2016dqh}.
This lifting to dS space is different from the KKLT scenario \cite{Kachru:2003aw}. 
Moreover, this model is valid in the weakly coupled regime $g_s=1/\Re S\ll 1$, where string loops can be safely neglected.
The K\"ahler uplift model has been well studied \cite{Westphal:2005yz,deAlwis:2011dp,Sumitomo:2012vx}, and so has 
 the racetrack model \cite{Krasnikov:1987jj,Taylor:1990wr,Denef:2004dm}. They are merged into the model where all parameters are replaced by flux parameters to be scanned over \cite{Sumitomo:2013vla}.

The superpotential $W_0 (U_i, S)$ and its supersymmetric solutions ($D_{U_i}W_0=D_SW_0=0$) have been studied in some detail \cite{Rummel:2011cd,Sumitomo:2012vx}. Here we simply state that $W_0(U_i,S,\phi)$ takes some value $\calw_0\equiv \mathcal{W}_0 (U_i, S,\phi)= W_0(U_i,S, \phi)|_{sol}$ after the equations have been solved. It turns out that the solved value of $\calw_0$ varies little by the K\"ahler uplift, i.e., solving the equation for $T$. Our goal here is to show that $\mathcal{W}_0$ is expected to be exponentially small. Here we closely follow Ref.\cite{Sumitomo:2013vla}, where more details can be found. 
Validity of a number of approximations taken can also be found in Ref.\cite{Rummel:2011cd,Sumitomo:2012vx}.

In the large volume region, $\Re T\gg1$, the resulting potential may be approximated to, with $T=t + i \tau$,
   \begin{align}
   \nonumber  &V \simeq \left(-{a^3 A \, \mathcal{W}_0\,  \over 2}\right) \lambda (x,y), \\
    & \lambda (x,y) =  - {e^{-x} \over x^2} \cos y - {\beta \over z} {e^{-\beta x} \over x^2} \cos (\beta y) + {\hat{C} \over x^{9/2}}, \qquad
  {\hat{C}} = -{3 a^{3/2} \calw_0 \, \xi \over 32 \sqrt{2} A},  \label{approxpot}
   \end{align}
 with $x = a t$, $y=a \tau$, $z = A/B$, $\beta = b/a=N_1/N_2 >1$. 
Extrema can be found by imposing $\partial_t V = \partial_{\tau} V = 0$, where the latter is immediately satisfied for $y=0$ (extrema with $y\neq0$ are not minima, see Ref.\cite{Sumitomo:2013vla}), while the former yields
 \begin{align} \frac{1}{z}=\frac{e^{\beta-1}\big(9\hat C e^{x}-2x^{5/2}(2+x)\big)}{2x^{5/2}\beta(2+\beta x)} \,,\label{zsol}\end{align}
where the $T$-dependence of $\mathcal{W}_0$ in the K\"ahler uplift is negligible \cite{Rummel:2011cd}. Plugging \eqref{zsol} into the Hessian (mass squared) components, and recasting the result in terms of $\lambda (x,y)$, we find $\partial _x \partial_y\lambda |_{\rm ext}  = 0$ and 
   \begin{align}
   \label{stability}
    &\partial_x^2 \lambda |_{\rm ext} \simeq  e^{-x} \left({\beta-1 \over x^2} - {5(\beta-1) \over 2 x^3} \right) + \lambda \left(-{9 \beta \over 2 x} - {9 \over 2 x^2} \right)  + \cdots = m_x^2 \ge 0 , \\ \nonumber
   &\partial_y^2 \lambda|_{\rm ext} \simeq e^{-x} \left(-{\beta-1 \over x^2} + {5 (\beta-1) \over 2 x^3} \right)+ \lambda \left({9 \beta \over 2 x} + {45 \over 4 x^2} \right) + \cdots = - m_x^2 +  \lambda \left({27 \over 4 x^2} \right) + \cdots  \ge 0. 
   \end{align}
So the stability condition (positive mass squared for both $x=at$ and $y=a\tau$ at the extremum) puts a strong constraint on the value of $\lambda=- 2 V|_{\rm ext}/ a^3 A \mathcal{W}_0 \ge 0$.
Requiring both of them to be positive (hence the extremum is a minimum) gives $ 0< \lambda_{\rm min} \lesssim \lambda \lesssim \lambda_{\rm max}$, or more precisely
\begin{equation}
 \begin{split}
e^{-x}{2 (\beta-1) \over 9 \beta x} \left(1 - {5 (\beta +1) \over 2 \beta x} \cdots \right)\lesssim \lambda \lesssim e^{-x}{2 (\beta-1) \over 9 \beta x}  \left(1 - {5(\beta + 1) \over 2 \beta x } + {3 \over 2 \beta x } \cdots \right) \,,
 \end{split}
 \label{leading lambda}
\end{equation}
and $A  \mathcal{W}_0 <0$. Here we have in mind $\mathcal{W}_0 >0$ and $A<0$, $\beta \gtrsim 1$ and $x \sim {\cal O}(100)$ respectively. So we see that a positive but small $\Lambda$ is guaranteed together with the large volume ${\cal V}$ and $\beta \gtrsim 1$.
For large $x$, $\lambda_{\rm min} \rightarrow \lambda_{\rm max}$ so at leading order, 
  \begin{align} &  \lambda \simeq e^{-x}\frac{2(\beta-1)}{9\beta x}, \quad \quad {\hat C} \simeq e^{-x} \frac{2(\beta-1)}{9\beta} x^{7/2} \label{CMM} \end{align} 
 and therefore $\Lambda$ approaches an exponentially small positive value at the large volume ($x \rightarrow \infty$) limit.  Using $\hat C$ \eqref{approxpot}, one can show that in this limit,
  \begin{align} \label{W0sol} \calw_0\sim-\frac{64\sqrt{2}A(\beta-1)}{27a^{3/2}\beta\xi}e^{-x}x^{7/2} \,,\end{align}
where the $x^{7/2}$ term is crucial to satisfy the assumption $\frac{|A|e^{-x}}{|\calw_0|}\ll1$. 
Using \eqref{CMM} and \eqref{W0sol}, we can easily obtain,
  \begin{align} \label{Key} \Lambda \simeq\frac{64\sqrt{2}a^{3/2}A^2(\beta-1)^2}{243 \beta^2\xi} e^{-2x}x^{5/2} \simeq \frac{3 \xi\calw_0^2}{4 (2t)^{9/2}} \,. \end{align}
As it might be expected from the exponential terms in \eqref{Key} and \eqref{W0sol}, the bigger the volume modulus, the smaller $\calw_0$ and  $\Lambda$ have to be in order to find a solution. Note that $\xi=0$ implies $\Lambda=0$, a property of the no-scale structure in supergravity.

As explained in Ref.\cite{Sumitomo:2013vla}, we can analyze the probability distribution of the cosmological constant, $P(\Lambda)$. 
After randomizing $A$, $B$ and $\calw_0$, we collect all the classically stable solutions and find that the probability distribution $P(\Lambda)$ for small positive $\Lambda$ is approximately given by \cite{Sumitomo:2013vla},
  \begin{align} P(\Lambda) \stackrel{\Lambda \rightarrow 0}{\sim} {243 \beta^{1/2} \over 16 (\beta-1)} {1 \over \Lambda^{\beta+1 \over 2 \beta}  (-\ln \Lambda)^{5/2}} \,.\label{asymptotics of PDF} \end{align}
So for $\beta \gtrsim 1$, we see that the diverging behavior of the properly normalized $P(\Lambda)$ is very peaked as  $\Lambda \rightarrow 0$.
We see that the expected value of $\Lambda$ is very sensitive to the value of $\beta$. Due to tadpole-cancellation and other constraints in F-theory, we expect the value of $N_{max}$ to be bounded. Ref.\cite{Louis:2012nb} finds that $N_{max}$ can easily exceed a hundred. 
In principle, we should also scan through all allowed values of $N$ in the $SU(N)$ gauge groups, i.e., $N=2,3,4,... N_{max}$. It turns out that the divergence of $P(\Lambda)$ at $\Lambda=0$ is dominated by the most divergent term, i.e., the smallest $\beta$, or $\beta=N_{max}/(N_{max}-1)$. For simplicity, we shall simply use the smallest allowed $\beta$ to perform our estimates.

It is informative to introduce $\Lambda_Y$, defined as $\int_0^{\Lambda_Y} d\Lambda \, P(\Lambda) = Y\%$. Here, $\Lambda_{50}$ is the median. It is interesting that one can find very simple formulae for $\Lambda_{10}$ and $\Lambda_{50}$ as a function of $N_{max}$,
   \begin{align} \Lambda_{10} \simeq 10^{1.57-1.91 N_{max}}, \quad \Lambda_{50} \simeq 10^{-2.61-0.59 N_{max}} \end{align}
when $N_{max}$ is large. 
In Table \ref{Tab1} and \ref{Tab2}, we present two cases, namely $\Lambda_{10}, \Lambda_{50}$ matching the observed $\Lambda \sim 10^{-122}$. 

\begin{table}[h!] 
\begin{center}
$\begin{array}{cccccccc}
\hline
  N_{max} & \Lambda_{10} & \Lambda_{50}  & \langle\Lambda\rangle  \\
\hline
  65 &  0.263 \times 10^{-122}  &  1.1 \times 10^{-41} & 6.11 \times 10^{-8}  \\
  202 & 5.62 \times 10^{-385} &  1.62 \times 10^{-122} & 3.79 \times 10^{-9}  \\
\hline
\end{array}$
\end{center}
\caption{\label{Tab1} Estimates for $\Lambda$ in units of $M_P$, where $\xi \simeq 10^{-3}$.}
\end{table}
\begin{table}[h!]  
\begin{center}
$\begin{array}{cccccccc}
\hline
  N_{max} & \beta & x & t & z & \hat C & |\calw_0|  \\
\hline
  65 & 1.016 & 142 & 1470 & -0.114 & 2.53 \times 10^{-57} & 4.23 \times 10^{-53}  \\
  202 & 1.005 & 141 & 4530 & -0.504 & 2.49 \times 10^{-57} & 1.24 \times 10^{-51}  \\
\hline
\end{array}$
\end{center}
\caption{\label{Tab2} Estimates for other parameters and modulus values when $\Lambda_{10}$ or $\Lambda_{50} \simeq 10^{-122}$, where $\xi \simeq 10^{-3}$.}
\end{table}
Setting the median $\Lambda_{50}$ equal to the observed $\Lambda \sim 10^{-122}M_P^4$, and recalling that the superpotential has mass dimension 3, Eq.(\ref{Key}) gives $|\calw_0 (U_i,S,\phi)| \simeq 10^{-51} M_P^3$ yielding a new mass scale
 \begin{align} {\bf m} = |\calw_0|^{1/3} \sim 10^{-17} M_P \sim 100 \,{\rm GeV} \, \label{Key2} \end{align}
 If we match the observed $\Lambda$ to $\Lambda_{10}$, that is, there is only a $10 \%$ probability that $\Lambda$ has a value smaller or equal to the observed value, we find that ${\bf m}$ drops by less than one order of magnitude.
 


\section{The Higgs-Like Sector}
\label{sec:higgs}

Perhaps the simplest way to implement a Higgs-like field in the effective theory for the racetrack K\"ahler uplift model  \cite{Sumitomo:2013vla} is via D3-brane separation. Let us briefly review the setup we have in mind.  Dynamical flux compactification introduces warp geometry due to branes, O-planes and background fluxes\cite{Giddings:2001yu}. 
That is, in Einstein frame, 
$$ds_{10}^2=e^{2w(y)}g_{\mu\nu}dx^\mu dx^\nu+e^{-2w(y)}\tilde g_{mn}(y)dy^mdy^n \,,$$ where $e^{-4w(y)}$ is the warp factor, the 4-dimensional metric $g_{\mu\nu}$ is either Minkowski, AdS or dS, while the 6-dimensional metric $\tilde g_{mn}$ is the underlying CY-like metric. 
A realistic picture envisions a bulk with warped throats attached to it. A D3-brane tends to sit at the bottom of a warped throat, the geometry of which may be described by a deformed conifold.  
Let the complexified D3-brane position for the $I$-th D3-brane be $Z_{I}^i$ ($i=1,2,3$), where we choose the co-ordinate where $Z^i=0$ at the tip of a particular conifold. However, because of the deformation, the bottom of that deformed (or resolved) conifold ends at $r_0$ (in the $\tilde g$ metric), and so all D3 positions must be $|Z_I^i| \ge r_0$. 
While the brane positions are good K\"ahler moduli for the effective field theory, the presence of D3-branes leads a redefinition of both the K\"ahler coordinate $T$ and the term $K_{\rm K}$ in the K\"ahler potential \eqref{LVS}. 
Using  $\Phi^i={Z^i}/{2\pi\alpha'}$  and setting momentarily $\xi=0$, we have
\cite{Martucci:2016pzt,Cownden:2016hpf,Chen:2008au}
\begin{align}
\label{TwKw}
&T=t+i\tau+\frac{\alpha'}{2}\sum_{I\in D3}k(\Phi_I,\bar \Phi _I)
\simeq 
t+i\tau+\frac{\alpha'}{2}\sum_{I\in D3} \Phi_I^i \bar \Phi_I^i\,,\\ \nonumber
& K = -3 \ln \left[T+ {\bar T} - \alpha' \sum_{I\in D3}k(\Phi_I,\bar \Phi _I) \right] \simeq - 3\ln (T+{\bar T}) +  3\alpha' \sum_{I\in D3}\frac{\Phi_I^i \bar \Phi_I^i}{T+ \bar T} + \cdots \,.
\end{align}
where the sum over $i$ is implied, and $k(\Phi,\bar \Phi)$ is the ``little'' K\"ahler potential of the underlying internal metric $\tilde g$ which we approximated to $\delta_{i\bar j}\Phi^i\bar\Phi^{\bar j}$. The expansion of the K\"ahler potential \eqref{TwKw} is valid in the regime where for each $I$-th brane
\begin{align}
\label{smallKH}
\frac{3\alpha' |\Phi_I|^2}{2t} \ll1 \,.
\end{align}
This situation is indeed realized in the large volume limit, $t\gg1$, with the branes sitting at the bottom $r_0$ of a warped throat, where $\alpha' |\Phi|^2 \simeq (2\pi)^{-2} \alpha'^{-1} r_0^2 \sim r_0^2/R^2$ is minimum (we used the fact that the size of the throat $R$ is typically of string scale $R\sim \alpha'^{1/2} \gg r_0$). In particular, $\alpha' |\Phi|^2 \sim 10^{-26}$ if $\Phi\sim 100$ GeV and $\alpha'\sim$ GUT scale). To leading approximation we can thus treat \eqref{smallKH} as a perturbation, alongside the $\alpha'$ correction $\xi$.

Now, consider two branes $I=1,2$ and define $\phi^i=(\Phi_1^i-\Phi_2^i)/{2},\varphi^i=(\Phi_1^i+\Phi_2^i)/{2}$. Since these are linear combinations of $\Phi_1^i,\Phi_2^i$, they are good K\"ahler coordinates. $\phi^i$ corresponds to the D3 separation and we identify one of its directions as our Higgs-like field $\phi$. Focusing only on $\phi$ for simplicity and recovering the $\xi$ term, the K\"ahler potential \eqref{TwKw} becomes 
$$-2 \log \bigg[(T+\bar T)^{3/2} + \frac{\xi}{2} \bigg] +  \frac{3 \alpha' |\phi|^2}{T + \bar T} \,,$$
and so 
\begin{align} \label{KH} K_H = \frac{3 \alpha' |\phi|^2}{T + \bar T} \,.\end{align}
is the Higgs-like $K_H$ in \eqref{LVS} which is relevant for us.
The picture is complete by choosing the following superpotential for $\phi$:
\begin{align} \label{W0p} W_0(\phi)= c_{\phi}
+ \mu\phi^2+\rho\phi^3 \, \end{align}
where $c_{\phi}$, $\mu$ and $\rho$ are independent flux parameters (or functions thereof) to be scanned over. 

Let us now insert this $W_0(\phi)$ \eqref{W0p} and $K_H(\phi)$ \eqref{KH} into the F-term potential $V$ \eqref{LVS} and compute $m_\phi$ and $\Lambda$. First, $S,U_i,\phi$ are stabilized supersymmetrically, $D_{S}W=D_{U_i}W=D_{\phi}W= 0$, and this is followed by the stabilization of $T$ via $\xi,W_{NP}$. Due to \eqref{smallKH}, to leading order $D_{\phi}W\simeq \de_{\phi}W_0(\phi)$, and this has two solutions:
in the absence of SSB, with $\phi=0$, or in case of spontaneous symmetry breaking (SSB), with $\phi=-{2\mu}/{3\rho}\neq0$ (at leading order). 

When $\phi=0$, it is clear that the analysis of the vacuum energy proceeds exactly as in the previous discussion. Indeed, since $t,\tau$ are much lighter than $S,U_i,\phi$ (whose masses are determined by fluxes), one can focus the attention on the effective theory in $T$, treating $S,U_i,\phi$ as already stabilized to their vevs. The only change with respect to the previous discussion is  $ c_1 \to c= c_1+c_{\phi}$ in
$\calw_0(U_i,S, \phi)$.
On the other hand, in order to find the order of magnitude of $m_\phi$, we can totally neglect $\alpha'$ and non-perturbative corrections. Using \eqref{KH}, the computation of the mass matrix gives at leading order (after canonicalising kinetic terms)
  \begin{align} \label{masses} m_{\phi}^2 & \simeq \frac{2\mu^2}{9 t}   \,. \end{align}
Taking into account $\xi,W_{NP}$ will confer a small mass to $t,\tau$ while shifting all other mass values by negligible amounts. The moduli masses $m_t$ and $m_{\tau}$ are typically much lighter than $\phi$ and are closer to $\Lambda/M_P^2$ \cite{Tye:2016jzi}. Without fine-tuning, we see that $\mu$ sets the electroweak-like scale.

In the case of SSB, $\phi\simeq-{2\mu}/{3\rho}\in\reals$, the situation gets a bit more involved.  It turns out that, to a good approximation, $m_{\phi}^2$ is still given by \eqref{masses}.\footnote{In order to obtain \eqref{masses}, we again neglect $\xi,W_{NP}$, and expand the mass matrix using \eqref{smallKH}. Moreover, we neglect off-diagonal kinetic terms, mixing $T,\phi$, which are sub-leading with respect to diagonal terms.  In this way, the canonicalisation proceeds as in the $\phi=0$ case.}
The potential \eqref{LVS} is also shifted by the $K_H$ term \eqref{KH}, but the correction is negligibly small. Let us briefly illustrate how this works. We look for extrema and impose positivity of the Hessian to find, for $x\gg1$,
\begin{align}
\label{Vsol}
V \simeq
-\frac{Aa^3\calw_0}{2}\big(\lambda(x,y)+\frac{d \hat C}{x^4}\big) \,,\quad
{\hat C} \to e^{-x}x^{7/2} \frac{2(\beta -1)}{(8d\sqrt{x} +9)\beta}  \,,\quad
d=\frac{16\sqrt{2}\alpha'\mu^2}{9a^{1/2}\xi \rho^2}\,,
\end{align}
with $\calw_0 = \calw_0(U,S) + c_\phi + 4\mu^3/27 \rho^2$ and  
\begin{equation}
\frac{1}{z}=\frac{e^{x(\beta-1)}\big((9+8dx^{1/2})\hat C e^{x}-2x^{5/2}(2+x)\big)}{2x^{5/2}\beta(2+\beta x)}
\,,
\label{zsol2}
\end{equation}
We see that the new term gives a negligible contribution to \eqref{zsol2} if $d x^{1/2} \ll 1$, that is when $\phi\sim\frac{\mu}{\rho}\ll 10^{-2}M_s$. Therefore, in this regime, \eqref{Vsol} collapses to \eqref{CMM}, and $\Lambda$ is essentially the same as in the $\phi=0$ case.

In short, in both vacua with $\phi=0$ and $\phi\neq0$, the order of magnitude of $m_\phi$ is determined by $\mu$ \eqref{masses}. The only difference is in $\calw_0(\phi)$: $c_{\phi}$ for $\phi=0$, and $c_\phi + 4\mu^3/27 \rho^2$ for $\phi\neq0$. 
Note that smooth probability distributions $P(\calw_0(U,S)),P(\calw_0(\phi))$ imply a smooth distribution $P(\calw_0)$. In fact, $P(\calw_0)$ can remain smooth even if both $P(\calw_0(U,S))$ and $P(\calw_0(\phi))$ peak at zero. In the absence of fine-tuning, one expects all the flux parameters $c, c_2, b_i, d_i$ (each with mass dimension 3) as well as $\mu^3$ to have comparable values. 
Assuming smooth probability distributions for the flux parameters, $P(\calw_0)$ also has a smooth distribution \cite{Sumitomo:2012vx}.
In general, this $\calw_0$ has a wide range. However, setting $\Lambda_{50}$ equal to the observed value, ${\bf m}^3 = \calw_0 \sim 10^{-51}M_P^3$ is required to yield meta-stable solutions.

In the SSB case, there are 3 possible scenarios:  \\
(a) $m_\phi\sim{\bf m}$ if ${\bf m}^3\simeq\calw_0 (\phi) \simeq \calw_0 (U_i,S)$ or $\calw_0 (\phi) > \calw_0 (U_i,S)$;  \\
(b)  $m_\phi\gg{\bf m}$, if ${\bf m}^3 \ll \calw_0 (\phi) \simeq \calw_0 (U_i,S)$ (when two terms (almost) cancel each other; \\
(c) $m_\phi\lesssim{\bf m}$, if ${\bf m}^3 \simeq \calw_0 (U_i,S) \gtrsim \calw_0 (\phi)$. \\
Since we have to scan over all values of the flux parameters, we expect that, in the absence of fine-tuning,
$$ \mu \sim {\bf m} \simeq 10^2 \, {\rm GeV}$$
within some orders of magnitude.

In more realistic versions of the model, $c_\phi,\mu,\rho$ are expected to be functions of fluxes and can depend on $U_i$ and $S$ \cite{Camara:2003ku}, and so $\calw_0 (\phi)$ and $\calw_0 (U_i,S)$ are actually coupled. In general, coupling different sectors tend to render them to have comparable scales, so statistically, we expect that they have the same magnitude, $\calw_0 (\phi)\simeq\calw_0 (U_i,S)$. 
In this sense, having ${\bf m}^3\equiv \calw_0(U_i,S,\phi)\simeq\calw_0 (\phi) \simeq \calw_0 (U_i,S)$ would be the most likely (and statistically natural) scenario, yielding a natural explanation also for the EW scale $\mu$.

As an illustration, let us consider a $W(\phi)$ that depends on $S$. Since the dilaton $S$ dictates all couplings (closed string coupling goes like $1/S$ so open string coupling $\rho \propto 1/\sqrt{S}$),  let us consider the simple case where (ignoring some order one numerical factors), 
$$ W_0(\phi,S) = c_{\phi} + 4\mu^3/27 \rho^2 \simeq c_{\phi} + S\mu^3$$ 
where we have substituted in the vev for $\phi$. Now we can solve the supersymmetric equations for $U_i$ and $S$, where $W_0(U_i,S,\phi)$ is now given by $W_0(U_i,S)$ in Eq.(\ref{LVS}) with $c_1 \to c=c_1+ c_{\phi}$ and   
\begin{align} 
\label{cc2} 
c_2 \to {\hat c}_2=c_2 - \mu^3 \quad {\rm SSB} 
\end{align} 
and no change in $c_2$ in the absence of SSB. The supersymmetric solution of $W(U_i, S)$ has been solved for real flux parameters \cite{Rummel:2011cd,Sumitomo:2012vx}. For example, for the $h^{2,1}=2$ complex structure moduli case, one finds that 
\begin{align} 
\label{old2} 
\calw_0 \equiv \calw_0(U_i, S,\phi) = \frac{(c+s{\hat c}_2)(b_1-sd_1)(b_2 -sd_2)}{s^2d_1d_2-b_1b_2} \,,  
\end{align} 
$s=c/{\hat c}_2$ and $\phi \simeq - \mu \sqrt{c/{\hat c}_2}$. 
Here $s=\Re S=c/{\hat c}_2 >1$ to stay in the weak coupling approximation. To satisfy Eq.(\ref{Key2}), we can take $\calw_0 \sim c$. Since couplings in the standard model is small but not vanishingly small, $s \gtrsim 1$, which implies $\calw_0 \sim {\hat c}_2$. Without fine-tuning, Eq.\eqref{cc2} suggests $\mu \sim {\bf m}$ and this combined with Eq.(\ref{Key2}) yields Eq.(\ref{L2}). 
Again, the uncertainty of $\mu$ is hard to estimate: $\mu \ll {\bf m}$ if $c \gg \mu^3$, and if $(b_i-sd_i) \to 0$, $c$ and ${\hat c}_2$ and so $\mu^3$ can be much bigger, i.e., $\mu \gg {\bf m}$. For $h^{2,1}>2$,  $\calw_0(U_i, S,\phi) \propto  (c+s{\hat c}_2)\Pi_i (b_i-sd_i)$, and $s \to c/{\hat c}_2$ if any of the factor $(b_i-sd_i) \to 0$. For more non-trivial couplings within $W_0(U_i, S, \phi)$, the analysis becomes more complicated. However, to get a better determination of $\mu$ with respect to ${\bf m}$, we need to determine the explicit functional forms of the flux parameters and their dependence on the moduli.

\section{Discussions and Summary}

It is shown that a very large fraction of the classically stable de Sitter vacua have an exponentially small positive $\Lambda$ in this racetrack K\"ahler uplift model \cite{Sumitomo:2013vla}, so it is likely that our universe ends in a vacuum with an exponentially small positive $\Lambda$.
Since this property is probabilistic, we cannot determine the precise value of $\Lambda$, but if we let the median $\Lambda_{50}$ to match the observed $\Lambda$, we find that a mass scale of ${\bf m} \sim 100$ GeV emerges. While the dilaton $S$, the complex structure moduli $U_i$ and K\"ahler modulus $T$ are all closed string modes, a Higgs-like scalar open string mode $\phi$ is introduced so its mass scale matches this intermediate scale, $\mu \sim {\bf m} \sim 100$ GeV. That is, we show for the first time how the electroweak and the cosmological constant scales are related \emph{dynamically}. 
It should be intuitively clear that replacing $\phi$ by a more realistic Higgs doublet (or two Higgs doublet) will not change this result by more than a few orders of magnitude.

Due to the relevance of this result, and especially in light of the recent conjectural concerns raised against dS solutions in string theory, let us make few remarks here.
\begin{itemize}
\item \underline{RKU and dS vacua} In an examination of the racetrack K\"ahler uplift (RKU) model \cite{Andriolo:2019gcb}, we note that only tiny ranges of some of the flux parameters lead to classically stable de Sitter vacua. That is, large patches of the flux landscape have no meta-stable de Sitter solution. In particular, for $N_{max} \gg 2$, only a tiny range of around a very small $\calw_0$ yields solutions. It is clear that without a sufficiently dense flux landscape, dS solutions may be not possible at all. See below for more discussions on this point.

In the very small patch of flux landscape with de Sitter vacua, we find that the probability distribution $P(\Lambda)$ sharply peaks at $\Lambda=0$ so $\Lambda$ is typically exponentially small, rendering the observed $\Lambda$ to be natural. Precisely because of the smallness of this ``de Sitter'' patch in the flux landscape, the observed $\Lambda$ requires that the superpotential $\calw_0$ of the standard or any other model to have a mass scale of $10^2$ GeV. It remains to be seen whether this is only a numerological accident or contains deep physical implications. 

\item \underline{Rolling to dS vacua} If the patch with a de Sitter solution in the flux landscape is so small, one may argue that the chance to end in such a patch is highly unlikely. That is, a random choice of flux parameters in the string landscape will result in a run-away solution \cite{Sethi:2017phn}. This may well be the case. A better understanding of the landscape is necessary before one can address this issue. However, we believe that the history of our universe may make this less unlikely.

Since the universe probably started from an inflationary epoch, which had a large $\Lambda$, rolling down to lower values of $\Lambda$ towards $\Lambda \le 0$ must pass through regions with small $\Lambda>0$. With the string scale close to the GUT scale, the sharp peaking of $P(\Lambda)$ at $\Lambda=0$ implies that there are exponentially many vacua with an exponentially small positive $\Lambda$; so rolling towards $\Lambda \le 0$ may be intercepted by one of the exponentially many meta-stable de Sitter vacua, before the universe can reach the vastly larger region in field space with $\Lambda \le 0$. So ending in a meta-stable de Sitter vacuum may not be that unlikely.

\item \underline{RKU and its sisters} The racetrack K\"ahler uplift model \cite{Sumitomo:2013vla} belongs to the same group of stringy-inspired 4d effective field theory as the KKLT model \cite{Kachru:2003aw}, and the K\"ahler uplift (KU) \cite{Westphal:2005yz,Rummel:2011cd}, of which it is a modification. In particular, the addition of a second non-perturbative piece $Be^{-bT}$ with respect to the original KU, makes a big difference. In fact, while dS vacua in the KU model are found where $\Lambda$ needs not be small, the presence of two more parameters $b,B$ open up the possibility for dS vacua at larger volume (e.g., $t\sim\calo(1000)$) and an exponentially small $\Lambda$. This justifies the use of the large volume limit and the fact that we neglect higher $\alpha'$ corrections. Moreover, since $\Lambda$ is exponentially suppressed with $t$, such vacua have smaller $\Lambda$. This is reflected in the probability distribution $P(\Lambda)$ which is very different from the original KU model \cite{Sumitomo:2012vx}. This same qualitative behavior persists if multiple NP terms are added into the race-track superpotential, as shown in \cite{Andriolo:2019gcb}. 

The (R)KU model shares also some similarities with the KKLT model, albeit being different in many aspects. As in the KKLT model, before considering the uplift the moduli stabilisation is achieved via NP effects. However, while in KKLT the AdS space thus constructed is supersymmetric, in this model supersymmetry is broken \cite{Sumitomo:2013vla}. The second is the uplift method. While KKLT uses an anti-D3 brane inside a warped throat, and the stability of this configuration is still under discussion, (R)KU simply uses the first and well-known $\alpha'$ correction to the K\"ahler potential \cite{Becker:2002nn}. KKLT and RKU also share the same need for an extremely small value of the on-shell superpotential, $\calw_0$. For instance, in our case, we need $\calw_0\sim 10^{-51} M_p^3$. It is hard to tell whether string theory allows for such small $\calw_0$, and a bound is still lacking. We hope to learn something from explicit numerical construction of flux vacua, see Ref.~\cite{Cole:2019enn} and references therein.

\item \underline{$P(\Lambda)$, radiative corrections and soft/D-terms} Our analysis is purely classical in the string coupling. This is because we assume to work in the weakly coupled regime $g_s\ll1$, where string loops can be safely neglected. However, the solutions we found can have $g_s\lesssim 0.8$ and one may worry that radiative corrections are relevant and spoil the exponentially small $\Lambda$. We can limit ourselves to smaller values of $g_s$ without spoiling the exponentially small $\Lambda$ preference. This possibility may also be checked in the simpler setup of a $\phi^3/\phi^4$ toy model, where classically $P(\Lambda)$ peaks. it was found that corrections (radiative or higher orders $\phi^5,\phi^6$) do not change the peaking behaviour of $P(\Lambda)$ \cite{Tye:2016jzi}. This is due to two facts. (1) Any convoluted function of randomised (flux) parameters $a_i$ as $\Lambda$ presents a peaking behaviour in its distribution, independently on the parameter distributions $P_i(a_i)$ \cite{Sumitomo:2012wa}. 
(2) We swipe through very dense flux values, and so the range of $\Lambda$ values before/after corrections is the same. The result is that if $\Lambda$ is complicated enough classically, corrections will not spoil the peaking behaviour of $P(\Lambda)$, but may rather change its details, e.g., shifting the median value $\Lambda_{50}$ by few orders of magnitude.

These considerations are expected to hold in general, and so also in any string inspired EFT as our model. In particular, the same reasoning can be applied for $\phi$ soft terms and D-terms, which we neglected in our model, but are expected to be present. In fact, the Higgs-like field $\phi$ is coupled to the other moduli sectors and, as such, the resulting $\Lambda$ will still satisfy point (1) above. But what about the relation between $\Lambda$ and ${\bf m}$? Notice that the quadratic term in the Higgs potential can come from the superpotential, while the quartic term may come from a D-term. Since the quadratic (i.e., mass) term is the term that sets the mass scale $\mu$ in the electroweak model, it is reasonable to link $\mu$ directly to $\bf m$. The remarkable result relating $\Lambda$ and ${\bf m}$ will then still hold. It is however important to find out how a D-term (or some other term) for the electroweak model couples to the moduli in the F-term $V$. This should put a tight constraint on the possible origin of such terms.

As a side remark, note that radiative corrections are of order of Higgs mass scale if supersymmetry breaking is of that order of magnitude.  That is, supersymmetry renders the mass hierarchy technically natural. However, it does not explain why the tree-level Higgs mass is of order of ${\bf m} \sim 100$ GeV, so it is still not natural. Here we provide a dynamical way within the string landscape to understand how the Higgs mass is natural, not only technically natural.

\item \underline{A ``dense discretuum'' of fluxes} From the above discussion, the importance of having dense enough flux values should be clear. It this always the case in string theory?

In string theory, the fluxes take discrete values. For a particular set of discrete flux values that yields a meta-stable vacuum with a specific $\Lambda \ge 0$ value around the observed dark energy value, Ref.\cite{Bousso:2000xa} proposes that neighboring meta-stable vacuua (i.e., those vacua with $\Lambda$'s closest to the specific $\Lambda$, with difference $\Delta \Lambda$) should be close enough so that $|\Delta \Lambda| \le |\Lambda|$. If true, the observed $\Lambda$ presumably can be reached. Ref.\cite{Bousso:2000xa} also estimates that this can be achieved if there are a dozen or more flux parameters. In that case, the $\Lambda$ values form a ``dense discretuum" and it is reasonable to allow a discrete flux value to be treated as a continuous parameter without qualitatively changing the estimates. In the model (\ref{LVS}) studied here, there are the parameters $A$ and $B$ in the non-perturbative terms as well as the terms in the superpotential $W_0(U_i,S, \phi)$ involving the complex structure moduli $U_i$, the dilaton $S$ and Higgs like fields $\phi$. Introducing more complex structure moduli and/or non-geometric couplings among them makes little difference to the overall picture, but will introduce more flux parameters so the number of flux parameters entering here can easily be dozens. The resulting dense discretuum yields a close-to-continuous $\calw_0(U_i, S, \phi)$, which in turn yields a dense discretuum for $\Lambda$. So it is reasonable to assume a dense discretuum in our analysis.

To simplify the analysis and because of a lack of better knowledge of the string theory properties, we treat each parameter as a flux parameter with a flat probability distribution. In the actual case, some parameters (e.g., $A$ and $B$) are known to be functions of flux parameters and other fields, so their probability distribution may not be flat. String theory studies suggest that there is a preference for small values for $A$ and $B$. In this case, the discrete spacing $\Delta A$ will be substantially smaller than the string scale, and we expect $P(A)$ and $P(B)$ to be smooth or peak at zero values. Ref.\cite{Sumitomo:2012vx} has investigated this issue and show that the qualitative behavior of $P(\Lambda)$ will not change much. Clearly a better understanding of the model will yield a more precise $P(\Lambda)$.  

Possible hindrances to the above reasoning may regard compactification details we neglected, e.g.~the tadpole cancellation. These additional conditions may act as selection rules, filtering allowed flux values. It remains to be understood whether this can pose a real threat to the dicretuum density.

\item \underline{No dS and other conjectures} It has been recently pointed out that not every EFT one can write down can be embedded into string theory, and criteria have been conjectured to distinguish the string Landscape from the Swampland \cite{Vafa:2005ui}. Such considerations kicked off a review of all stringy inspired 4d EFT's. Here we would like to clarify where our model stands with respect to these claims.

(a) \emph{no dS conjecture:} More than twenty years ago, Dine and Seiberg proposed that there is no stable de Sitter vacuum in string theory in the \emph{asymptotic} regime of weak coupling $g_s\to0$ ($s\to\infty$) \cite{Dine:1985he}. Our solutions, at weakly but finite coupling $g_s\sim\calo(10^{-1})$ ($s\sim\calo(1)$) are then untouched by this consideration. More recently, the rationale of \cite{Dine:1985he} has been extended to any modulus (e.g., the volume modulus $t$), and it has been conjectured that string theory does not allow for (meta-)stable dS vacua at all \cite{Agrawal:2018own,Ooguri:2018wrx}. This is an extrapolation from the asymptotic regime (e.g., $t\to\infty$), while explicit finite bounds are still absent. As before, our solutions at large but finite volume do not clash with the conjecture, and actually agree with it in the asymptotic $t\to\infty$ where we observe a runaway.

(b) \emph{SUSY breaking by fluxes:} Ref.~\cite{Sethi:2017phn} raises the concern that in case supersymmetry is broken by fluxes, i.e.~$\calw_0\neq0$ (as in our case), all EFT's \emph{built so far} are plagued by the same issue: truncation to any order in $\alpha'$ is inconsistent, the background is not static but rolling (with time), and therefore NP effects have to be re-considered. In other words, it is not correct to use the NP effects we are accustomed with (e.g., gaugino condensation on D7's \cite{Krasnikov:1987jj,Taylor:1990wr,Denef:2004dm}) to stop the runaway, since these are obtained from a computation on a static background. The point raised is a good one, but unfortunately it kills any room for manoeuvre. It would be first necessary to fully understand NP effect on non-static backgrounds, as started in \cite{Pimentel:2019otp}, and then apply it to string theory. In any case, it would be very surprising if the \emph{right} NP effects can never balance the runaway and create a dS vacuum. 

\item \underline{How about other scales in nature?} One is the inflaton scale in the inflationary universe scenario. It turns out that the scale of the inflaton potential is comparable to the string scale, so new scale needs to be generated. Recent data indicates that the brane inflation model, natural in string theory (where $\partial^2V<0$),  provides an excellent fit to the observation \cite{Akrami:2018odb}. 
Another is the fuzzy dark matter mass scale, given by a scalar field with mass $\sim10^{-22}$ eV.  It turns out that some string moduli \cite{Tye:2016jzi}, in particular axionic modes \cite{Hui:2016ltb}, will fit in nicely. In short, the string theory landscape with no explicit parameter except the string scale is a fruitful playground to explore nature. 
We believe that the search for the standard model of electroweak and QCD should only take place in regions of the landscape where $\Lambda>0$ is exponentially small.


\end{itemize}

In summary, both the K\"ahler uplift model and the racetrack model are well studied in string phenomenology, so putting them together is natural \cite{Sumitomo:2013vla}.  A patch of the string theory landscape, generated by treating all parameters not as free parameters but as flux parameters or functions of flux parameters that we scan over, is crucial in allowing the statistical preference for an exponentially small $\Lambda>0$ and the emergence of the electroweak scale without knowing the specific electroweak model. It will be very interesting to embed the known supersymmetric electroweak (phenomenological) model into this RKU framework.

 \section*{Acknowledgments} 
  We thank Michael Haack, Daniel Junghans, Luca Martucci, Toshifumi Noumi, Ander Retolaza, Sam Wong and especially Gianluca Zoccarato for useful discussions.


\bibliographystyle{utphys}
\bibliography{refs}

\providecommand{\href}[2]{#2}\begingroup\raggedright\begin{thebibliography}{10}

\bibitem{Sumitomo:2013vla}
Y.~Sumitomo, S.~H.~H. Tye and S.~S.~C. Wong,  {\em {Statistical Distribution of
  the Vacuum Energy Density in Racetrack K\"ahler Uplift Models in String
  Theory}}, JHEP {\bf 07} (2013) 052
[\href{http://www.arXiv.org/abs/1305.0753}{{\tt 1305.0753}}].

\bibitem{Hinshaw:2012aka}
{WMAP} Collaboration, G.~Hinshaw {\em et al.},  {\em {Nine-Year Wilkinson
  Microwave Anisotropy Probe (WMAP) Observations: Cosmological Parameter
  Results}}, Astrophys. J. Suppl. {\bf 208} (2013) 19
[\href{http://www.arXiv.org/abs/1212.5226}{{\tt 1212.5226}}].

\bibitem{Balasubramanian:2004uy}
V.~Balasubramanian and P.~Berglund,  {\em {Stringy corrections to Kahler
  potentials, SUSY breaking, and the cosmological constant problem}}, JHEP {\bf
  11} (2004) 085
[\href{http://www.arXiv.org/abs/hep-th/0408054}{{\tt hep-th/0408054}}].

\bibitem{Westphal:2006tn}
A.~Westphal,  {\em {de Sitter string vacua from Kahler uplifting}}, JHEP {\bf
  03} (2007) 102
[\href{http://www.arXiv.org/abs/hep-th/0611332}{{\tt hep-th/0611332}}].

\bibitem{Rummel:2011cd}
M.~Rummel and A.~Westphal,  {\em {A sufficient condition for de Sitter vacua in
  type IIB string theory}}, JHEP {\bf 01} (2012) 020
[\href{http://www.arXiv.org/abs/1107.2115}{{\tt 1107.2115}}].

\bibitem{deAlwis:2011dp}
S.~de~Alwis and K.~Givens,  {\em {Physical Vacua in IIB Compactifications with
  a Single Kaehler Modulus}}, JHEP {\bf 10} (2011) 109
[\href{http://www.arXiv.org/abs/1106.0759}{{\tt 1106.0759}}].

\bibitem{Sumitomo:2012vx}
Y.~Sumitomo and S.~H.~H. Tye,  {\em {A Stringy Mechanism for A Small
  Cosmological Constant - Multi-Moduli Cases -}}, JCAP {\bf 1302} (2013) 006
[\href{http://www.arXiv.org/abs/1209.5086}{{\tt 1209.5086}}].

\bibitem{Krasnikov:1987jj}
N.~V. Krasnikov,  {\em {On Supersymmetry Breaking in Superstring Theories}},
  Phys. Lett. {\bf B193} (1987)
37--40.

\bibitem{Taylor:1990wr}
T.~R. Taylor,  {\em {Dilaton, gaugino condensation and supersymmetry
  breaking}}, Phys. Lett. {\bf B252} (1990)
59--62.

\bibitem{Denef:2004dm}
F.~Denef, M.~R. Douglas and B.~Florea,  {\em {Building a better racetrack}},
  JHEP {\bf 06} (2004) 034
[\href{http://www.arXiv.org/abs/hep-th/0404257}{{\tt hep-th/0404257}}].

\bibitem{Bousso:2000xa}
R.~Bousso and J.~Polchinski,  {\em {Quantization of four form fluxes and
  dynamical neutralization of the cosmological constant}}, JHEP {\bf 06} (2000)
  006
[\href{http://www.arXiv.org/abs/hep-th/0004134}{{\tt hep-th/0004134}}].

\bibitem{Tye:2016jzi}
S.~H.~H. Tye and S.~S.~C. Wong,  {\em {Linking Light Scalar Modes with A Small
  Positive Cosmological Constant in String Theory}}, JHEP {\bf 06} (2017) 094
[\href{http://www.arXiv.org/abs/1611.05786}{{\tt 1611.05786}}].

\bibitem{Sumitomo:2012wa}
Y.~Sumitomo and S.~H.~H. Tye,  {\em {A Stringy Mechanism for A Small
  Cosmological Constant}}, JCAP {\bf 1208} (2012) 032
[\href{http://www.arXiv.org/abs/1204.5177}{{\tt 1204.5177}}].

\bibitem{Lust:2005dy}
D.~Lust, S.~Reffert, W.~Schulgin and S.~Stieberger,  {\em {Moduli stabilization
  in type IIB orientifolds (I): Orbifold limits}}, Nucl. Phys. {\bf B766}
  (2007) 68--149
[\href{http://www.arXiv.org/abs/hep-th/0506090}{{\tt hep-th/0506090}}].

\bibitem{Lust:2006zg}
D.~Lust, S.~Reffert, E.~Scheidegger, W.~Schulgin and S.~Stieberger,  {\em
  {Moduli Stabilization in Type IIB Orientifolds (II)}}, Nucl. Phys. {\bf B766}
  (2007) 178--231
[\href{http://www.arXiv.org/abs/hep-th/0609013}{{\tt hep-th/0609013}}].

\bibitem{Kachru:2003aw}
S.~Kachru, R.~Kallosh, A.~D. Linde and S.~P. Trivedi,  {\em {De Sitter vacua in
  string theory}}, Phys. Rev. {\bf D68} (2003) 046005
[\href{http://www.arXiv.org/abs/hep-th/0301240}{{\tt hep-th/0301240}}].

\bibitem{Becker:2002nn}
K.~Becker, M.~Becker, M.~Haack and J.~Louis,  {\em {Supersymmetry breaking and
  alpha-prime corrections to flux induced potentials}}, JHEP {\bf 06} (2002)
  060
[\href{http://www.arXiv.org/abs/hep-th/0204254}{{\tt hep-th/0204254}}].

\bibitem{Bonetti:2016dqh}
F.~Bonetti and M.~Weissenbacher,  {\em {The Euler characteristic correction to
  the Kähler potential — revisited}}, JHEP {\bf 01} (2017) 003
[\href{http://www.arXiv.org/abs/1608.01300}{{\tt 1608.01300}}].

\bibitem{Westphal:2005yz}
A.~Westphal,  {\em {Eternal inflation with alpha-prime-corrections}}, JCAP {\bf
  0511} (2005) 003
[\href{http://www.arXiv.org/abs/hep-th/0507079}{{\tt hep-th/0507079}}].

\bibitem{Louis:2012nb}
J.~Louis, M.~Rummel, R.~Valandro and A.~Westphal,  {\em {Building an explicit
  de Sitter}}, JHEP {\bf 10} (2012) 163
[\href{http://www.arXiv.org/abs/1208.3208}{{\tt 1208.3208}}].

\bibitem{Giddings:2001yu}
S.~B. Giddings, S.~Kachru and J.~Polchinski,  {\em {Hierarchies from fluxes in
  string compactifications}}, Phys. Rev. {\bf D66} (2002) 106006
[\href{http://www.arXiv.org/abs/hep-th/0105097}{{\tt hep-th/0105097}}].

\bibitem{Martucci:2016pzt}
L.~Martucci,  {\em {Warped Kähler potentials and fluxes}}, JHEP {\bf 01}
  (2017) 056
[\href{http://www.arXiv.org/abs/1610.02403}{{\tt 1610.02403}}].

\bibitem{Cownden:2016hpf}
B.~Cownden, A.~R. Frey, M.~C.~D. Marsh and B.~Underwood,  {\em {Dimensional
  Reduction for D3-brane Moduli}}, JHEP {\bf 12} (2016) 139
[\href{http://www.arXiv.org/abs/1609.05904}{{\tt 1609.05904}}].

\bibitem{Chen:2008au}
F.~Chen and H.~Firouzjahi,  {\em {Dynamics of D3-D7 Brane Inflation in
  Throats}}, JHEP {\bf 11} (2008) 017
[\href{http://www.arXiv.org/abs/0807.2817}{{\tt 0807.2817}}].

\bibitem{Camara:2003ku}
P.~G. Camara, L.~E. Ibanez and A.~M. Uranga,  {\em {Flux induced SUSY breaking
  soft terms}}, Nucl. Phys. {\bf B689} (2004) 195--242
[\href{http://www.arXiv.org/abs/hep-th/0311241}{{\tt hep-th/0311241}}].

\bibitem{Andriolo:2019gcb}
S.~Andriolo, S.~Y. Li and S.~H.~H. Tye,  {\em {String Landscape and Fermion
  Masses}},
\href{http://www.arXiv.org/abs/1902.06608}{{\tt 1902.06608}}.

\bibitem{Sethi:2017phn}
S.~Sethi,  {\em {Supersymmetry Breaking by Fluxes}}, JHEP {\bf 10} (2018) 022
[\href{http://www.arXiv.org/abs/1709.03554}{{\tt 1709.03554}}].

\bibitem{Cole:2019enn}
A.~Cole, A.~Schachner and G.~Shiu,  {\em {Searching the Landscape of Flux Vacua
  with Genetic Algorithms}},
\href{http://www.arXiv.org/abs/1907.10072}{{\tt 1907.10072}}.

\bibitem{Vafa:2005ui}
C.~Vafa,  {\em {The String landscape and the swampland}},
\href{http://www.arXiv.org/abs/hep-th/0509212}{{\tt hep-th/0509212}}.

\bibitem{Dine:1985he}
M.~Dine and N.~Seiberg,  {\em {Is the Superstring Weakly Coupled?}}, Phys.
  Lett. {\bf B162} (1985)
299--302.

\bibitem{Agrawal:2018own}
P.~Agrawal, G.~Obied, P.~J. Steinhardt and C.~Vafa,  {\em {On the Cosmological
  Implications of the String Swampland}}, Phys. Lett. {\bf B784} (2018)
  271--276
[\href{http://www.arXiv.org/abs/1806.09718}{{\tt 1806.09718}}].

\bibitem{Ooguri:2018wrx}
H.~Ooguri, E.~Palti, G.~Shiu and C.~Vafa,  {\em {Distance and de Sitter
  Conjectures on the Swampland}}, Phys. Lett. {\bf B788} (2019) 180--184
[\href{http://www.arXiv.org/abs/1810.05506}{{\tt 1810.05506}}].

\bibitem{Pimentel:2019otp}
G.~L. Pimentel and J.~Stout,  {\em {Real-Time Corrections to the Effective
  Potential}},
\href{http://www.arXiv.org/abs/1905.00219}{{\tt 1905.00219}}.

\bibitem{Akrami:2018odb}
{Planck} Collaboration, Y.~Akrami {\em et al.},  {\em {Planck 2018 results. X.
  Constraints on inflation}},
\href{http://www.arXiv.org/abs/1807.06211}{{\tt 1807.06211}}.

\bibitem{Hui:2016ltb}
L.~Hui, J.~P. Ostriker, S.~Tremaine and E.~Witten,  {\em {Ultralight scalars as
  cosmological dark matter}}, Phys. Rev. {\bf D95} (2017), no.~4, 043541
[\href{http://www.arXiv.org/abs/1610.08297}{{\tt 1610.08297}}].

\end{thebibliography}\endgroup

\end{document}